\documentclass[twocolumn,letter]{jpsj3}

\title{
Anisotropy of Magnetoresistance Hysteresis around the $\nu=2/3$ Quantum Hall State in Tilted Magnetic Field
}

\author{Kazuki Iwata\thanks{E-mail address:\,iwata@tfu-mail.tfu.ac.jp}, Masayuki Morino$^{1}$, Akira Fukuda$^{2}$, Norio Kumada$^{3}$, Zyun Francis Ezawa$^{4}$,\\
Yoshiro Hirayama$^{1}$, and Anju Sawada$^{5}$ 
}

\inst
{Graduate School of Science, Kyoto University, Kyoto, 606-8502\\
$^{1}$Graduate School of Science, Tohoku University, Sendai, 980-8578\\
$^{2}$Department of Physics, Hyogo College of Medicine, Nishinomiya 663-8501\\
$^{3}$NTT Basic Research Laboratories, NTT Corporation, Atsugi, 243-0198\\
$^{4}$Theoretical Physics Laboratory, RIKEN, Saitama, 351-0198\\
$^{5}$Research Center for Low Temperature and Materials Sciences, Kyoto University, Kyoto, 606-8501
}

\abst{
We present an anisotropy of the hysteretic transport around the spin transition point at the Landau level filling factor $\nu=2/3$ in a tilted magnetic field.
When the direction of the in-plane component of the magnetic field $B_{\parallel}$ is normal to the probe current $I$, a strong hysteretic transport due to the current-induced nuclear spin polarization occurs.
When $B_{\parallel}$ is parallel to $I$, on the other hand, the hysteresis almost disappears.
We also demonstrate that the nuclear spin-lattice relaxation rate $T_{1}^{-1}$ at the transition point increases with decreasing angle between the directions of $B_{\parallel}$ and $I$.
These results suggest that the morphology of electron spin domains around $\nu =2/3$ is affected by the current direction.
\\
\\
\sffamily {KEYWORDs: two-dimensional electron system, fractional quantum Hall effect, phase transition, domain structure, nuclear spin system, hyperfine interaction, nuclear spin-lattice relaxation rate}
}

\begin{document}
\maketitle

An incompressible quantum liquid state of the fractional quantum Hall (QH) effect occurs as a consequence of strong electron-electron interactions in high magnetic field and has continued to be one of the most absorbing subjects in solid state physics \cite{text}.
In addition, the spin degree of freedom provides a rich variety of quantum phenomena in fractional QH states.
Because of the small Zeeman energy of electrons in GaAs, ground states with spin configurations other than the fully polarized one are possible, and transitions between different spin states have been observed at various fractional filling factors $\nu $ \cite{Eisenstein_etc_PRB,Kukushkin _PRL,Kumada_PRL_PRB}.

At the spin transition point of $\nu=2/3$, an anomalous magnetoresistance $R_ {xx} $ peak with hysteretic transport has been observed \cite{Kumada_PRL_PRB, Kronmuller_PRL, Smet_Nature, Hashimoto_PRL1, Kraus_PRL,Stern_PRB}.
Resistively\,-\,detected nuclear magnetic resonance measurements have shown the involvement of the nuclear spin polarization in the origin of the $R_ {xx}$ peak \cite{Kronmuller_PRL,Hashimoto_PRL1,Kraus_PRL,Stern_PRB}.
At the transition point, two fractional QH states with different spin configurations degenerate energetically and an electronic domain structure is formed \cite{Verdene_NaturePhys,Stern_PRB}.
Because the domain wall between the two fractional QH states connects the edge channels with opposite current direction, the backscattering of electrons through the domain wall increases $R_{xx}$ \cite{Jungwirth_PRL_PRB}.
In addition, it is believed that when a current passes across a domain boundary, electron spins flip-flop scatter nuclear spins, inducing nuclear spin polarization \cite{Hashimoto_PRL1,Kraus_PRL,Stern_PRB}.
The current-induced nuclear spin polarization reacts on the electronic domain structure through the hyperfine interaction and changes the properties of the domain structure.
This mutual interplay between the domain structure and the current-induced nuclear spin polarization causes a long-time variation of $R_{xx}$ in proportion to the nuclear polarization \cite{Kronmuller_PRL,Hashimoto_PRL1}.
Moreover, hysteresis in $R_{xx}$, taken with upward and downward field sweeps, has been observed around the transition point, indicating that the degree of nuclear spin polarization depends on the sweep direction \cite{Kumada_PRL_PRB,Kraus_PRL,Stern_PRB}.
However, despite intensive works, details of the domain structure and the mechanism of the current-induced nuclear spin polarization are still unclear.

In this report, we investigate the hysteresis around $\nu =2/3$ in a tilted magnetic field to understand how the domain structure is developed by interplay between the domains and nuclear spin polarization.
We present the anisotropic behavior of the hysteresis against the relative directions of the in-plane magnetic field $B_{\parallel}$ and the probe current $I$.
When the direction of $I$ is orthogonal to that of $B_{\parallel}$, a large hysteresis appears.
When $I$ flows parallel to $B_{\parallel}$, on the other hand, the hysteresis almost disappears.
By changing the angle $\phi $ between the directions of $I$ and $B_\parallel $ continuously, we demonstrate that the hysteresis becomes weaker with decreasing $\phi$.
Moreover, we measure nuclear spin-lattice relaxation rate $T_{1}^{-1}$ at the spin phase transition point of $\nu=2/3$.
$T_{1}^{-1}$ increases gradually with decreasing $\phi$, as is consistent with the behavior of the hysteresis.
These results suggest that the morphology of the domain structure is affected by the $I$ direction and the properties of electron spin fluctuations in domain walls depend on the $B_{\parallel}$ direction.


The sample employed in this study consists of a 20-nm-wide GaAs/AlGaAs quantum well and has been processed into a 50-$\mu$m-wide Hall bar with a voltage probe distance of 180\,$\mu$m.
Electron density $n$ can be controlled by a gate.
The low-temperature mobility is $8.0 \times 10^{5}$\,cm$^{2}$/Vs at $n = 0.8 \times 10^{11}$\,cm$^{-2}$.
We measured $R_{xx}$ and Hall resistance $R_{xy}$ by a standard low-frequency AC lock-in technique with a frequency of 17.7\,Hz in a dilution refrigerator with a base temperature of $T=70$\,mK.
The magnetic field is swept slowly (0.13\,T/min) to observe the hysteresis \cite{Hashimoto_PRL1}.

To investigate the anisotropic transport with respect to the direction of $B_\parallel$, we developed a high-precision two-axis {\it in situ} goniometer \cite{Suzuki_cryogenics}, with which we can control not only $\phi$ (inset in Fig.\,1(b)) but also the azimuthal angle $\theta$ representing the ratio of $B_\parallel$ to the perpendicular field $B_\perp$, i.e., $\theta = \arctan (B_\parallel/B_\perp)$ [inset in Fig.\,1(a)].
These angles are calibrated by $R_{xy}$ of the sample and a conventional Hall element attached to the goniometer perpendicular to the sample.


\begin{figure}
\begin{center}
   \includegraphics[height=.24\textheight]{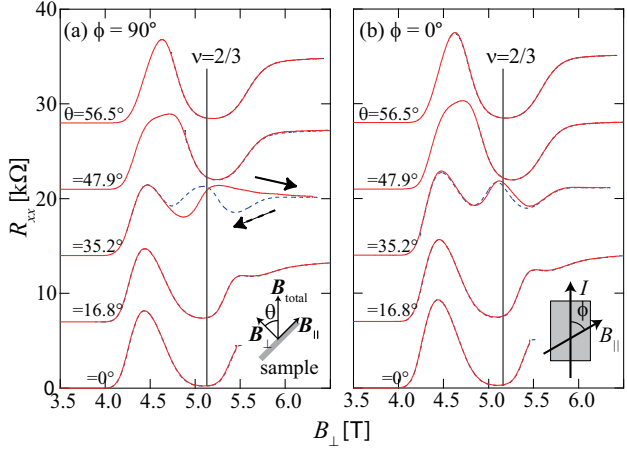}
\end{center}
\caption{(Color online) 
$R_{xx}$ as a function of $B_\perp$ for several values of $\theta$ around $\nu = 2/3$ at $n = 0.8 \times 10^{11}$\,cm$^{-2}$ and $T=70$\,mK.
$I$ passes perpendicular to $B_\parallel $ ($\phi =90^\circ $) in (a), and parallel to $B_\parallel$ ($\phi =0^\circ $) in (b).
The solid and dashed traces are obtained by sweeping the magnetic field upward and downward, respectively.
Traces are vertically offset by 7\,k$\Omega$ for clarity.
The vertical lines represent $B_\perp $ for just $\nu=2/3$.
The insets in (a) and (b) illustrate the definitions of $\theta$ and $\phi$, respectively.
}
\end{figure}
First, we demonstrate anisotropic transport around the $\nu = 2/3$ QH state under the tilted magnetic field. Figure\,1 shows $R_{xx}$ as a function of $B_{\perp}$ around $\nu = 2/3$ for several $\theta$ with $I=10$\,nA at $n = 0.8 \times 10^{11}$\,cm$^{-2}$. In Figs.\,1(a) and 1(b), $\phi$ is fixed at $90^{\circ}$ and $0^{\circ}$, respectively. The solid (dashed) traces represent $R_{xx}$ taken with upward (downward) sweep of the field. For the both values of $\phi$, as $\theta $ is increased from $\theta =0^{\circ}$, the well-developed $R_{xx}$ minimum at $\nu =2/3$ disappears at $\theta =35.2^{\circ}$ and reappears for larger $\theta$. At $\theta =35.2^{\circ}$, while a hysteresis between upward and downward field sweeps appears at $\phi=90^\circ $, it almost disappears at $\phi=0^\circ $.

 The disappearance and reappearance of the $R_{xx}$ minimum are known to be caused by the phase transition from the spin-unpolarized state for smaller $\theta $ to the spin-polarized state for larger $\theta$ \cite{text,Eisenstein_etc_PRB,Kukushkin_PRL,Kumada_PRL_PRB,Kraus_PRL,Stern_PRB}. At small and large $\theta$, $R_{xx}$ for $\phi=0^{\circ}$ and $90^{\circ}$ shows similar behavior, indicating that the relative direction between $I$ and $B_{\parallel}$ does not influence the stability of the fractional QH state or the dissipation of the electrons. However, at the transition angle $\theta =35.2^{\circ}$, there is a large difference in the appearance of the hysteresis, which is related to the current-induced nuclear spin polarization \cite{Kumada_PRL_PRB,Kraus_PRL,Stern_PRB}. At the transition angle, it has been argued that electronic domains are formed and the nuclear polarization is induced by the flip-flop process around domain walls \cite{Kumada_PRL_PRB,Hashimoto_PRL1,Kraus_PRL,Stern_PRB}. 
Therefore, the anisotropy of the hysteresis indicates that either the domain structure or the nuclear spin polarization or both are affected by the relative direction between $I$ and $B_{\parallel}$.
It is worth noting that since the spin transition in the $\nu =2/3$ QH state can be induced by changing $n$ \cite{Kumada_PRL_PRB,Stern_PRB}, we can investigate the hysteresis at a small angle, $\theta = 1.9^{\circ}$, with a higher density, $n = 1.2 \times 10^{11}$\,cm$^{-2}$. In this case, there is no anisotropy between $\phi=0^{\circ}$ and $90^{\circ}$, showing that the magnitude of $B_{\parallel}$ is important for inducing the anisotropy.

\begin{figure}
\begin{center}
   \includegraphics[height=.24\textheight]{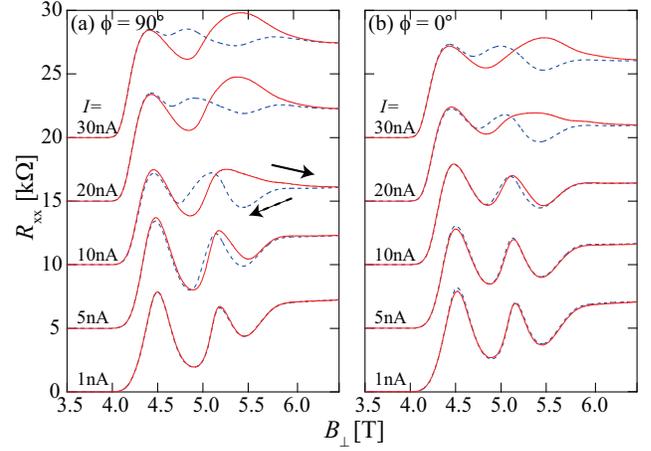}
\end{center}
\caption{(Color online) 
$R_{xx}$ as a function of $B_\perp $ for several values of $I$ for (a) $\phi = 0^{\circ}$ and (b) $\phi = 90^{\circ}$ at $T=70$\,mK and $\theta =35.2^{\circ}$.
The electron density is $n = 0.8 \times 10^{11}$\,cm$^{-2}$.
Arrows indicate the directions of the field sweeps.
Traces are vertically offset by 5\,k$\Omega$ for clarity.
}
\end{figure}

To elucidate the anisotropic behavior of the hysteresis, we measure the current dependence of the hysteresis while $\theta$ is fixed at the spin transition angle $\theta =35.2^{\circ}$.
Figures\,2(a) and 2(b) display $R_{xx}$ for several values of $I$ at $T=70$\,mK for $\phi =90^\circ $and $0^\circ$, respectively.
When the current is small ($I = 1$\,nA), the hysteresis does not occur for either angle.
As the current is increased to $I = 5$ and 10\,nA, the hysteresis emerges only at $\phi = 90^{\circ}$.
On increasing $I$ further to $I = 20$ and 30\,nA, the hysteresis appears at not only $\phi=90^{\circ}$ but also $\phi=0^{\circ}$.

The disappearance of the hysteresis at $I=1$\,nA is consistent with previous results \cite{Kraus_PRL,Stern_PRB}\,; at such a small current, the nuclear spin polarization is too small to drive the hysteresis.
For 5 $< I <$ 20\,nA, strong anisotropy appears, indicating that there is a large anisotropy of the magnitude of the nuclear polarization.
For larger current ($I>20$\,nA), as we will show below, the rate of nuclear spin pumping by the current is much faster than the nuclear spin relaxation rate and the strong hysteresis appears regardless of $\phi$.

We also investigate the $\phi$ dependence of the hysteresis at $\theta=35.2^{\circ}$.
Figure\,3 shows $R_{xx}$ as a function of $B_{\perp}$ at several $\phi$ at $T=70$\,mK.
We apply the current of $I=10$\,nA to investigate the anisotropy.
The strongest hysteresis appears at $\phi=90^{\circ}$, becomes weak gradually with decreasing $\phi$, and then almost disappears at $\phi=0^{\circ}$.
These results show that the anisotropy emerges when the rate of current-induced nuclear polarization is finite but relatively small.

\begin{figure}
\begin{center}
   \includegraphics[height=.30\textheight]{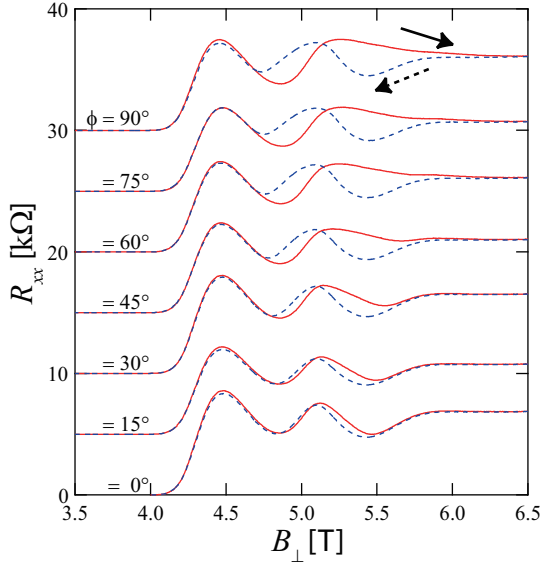}
\end{center}
\caption{(Color online) 
$R_{xx}$ vs $B_\perp$ for several $\phi$ at the transition angle ($\theta = 35.2^\circ$) with $I=10$\,nA.
The temperature and the electron density are $T=70$\,mK and $n = 0.8 \times 10^{11}$\,cm$^{-2}$, respectively.
Traces are vertically offset by 5\,k$\Omega$ for each curve.
Solid and dashed lines correspond to traces at upward and downward field sweeps, respectively.
As $\phi$ is decreased, the hysteresis becomes small and almost disappears at $\phi = 0^\circ$.
}
\end{figure}

To investigate the origin of the anisotropy, we study the nuclear spin-lattice relaxation rate $T_{1}^{-1}$ and the nuclear spin pumping rate.
Figure 4 (a) shows the time evolution of $R_ {xx}$ at the spin transition point at $\theta=35.2^{\circ}$.
When the large current ($I=10$\,nA) is applied, $R_{xx}$ increases with time because of pumping of the nuclear spin polarization.
We wait about 800\,s from the increment of the current and then reduce the current to 1\,nA, where $R_{xx}$ decreases with time to the value before the current pumping.
Figure\,4(b) shows log-scale plots of $\Delta R_{xx}$ defined as $R_{xx}(t)-R_{xx}(0)$ as a function of time after $I$ is reduced to $I=1$\,nA ($t > 850$ s).
Open and closed circles denote the data for $\phi=90^{\circ}$ and $0^{\circ}$, respectively.

\begin{figure}
\begin{center}
   \includegraphics[height=.25\textheight]{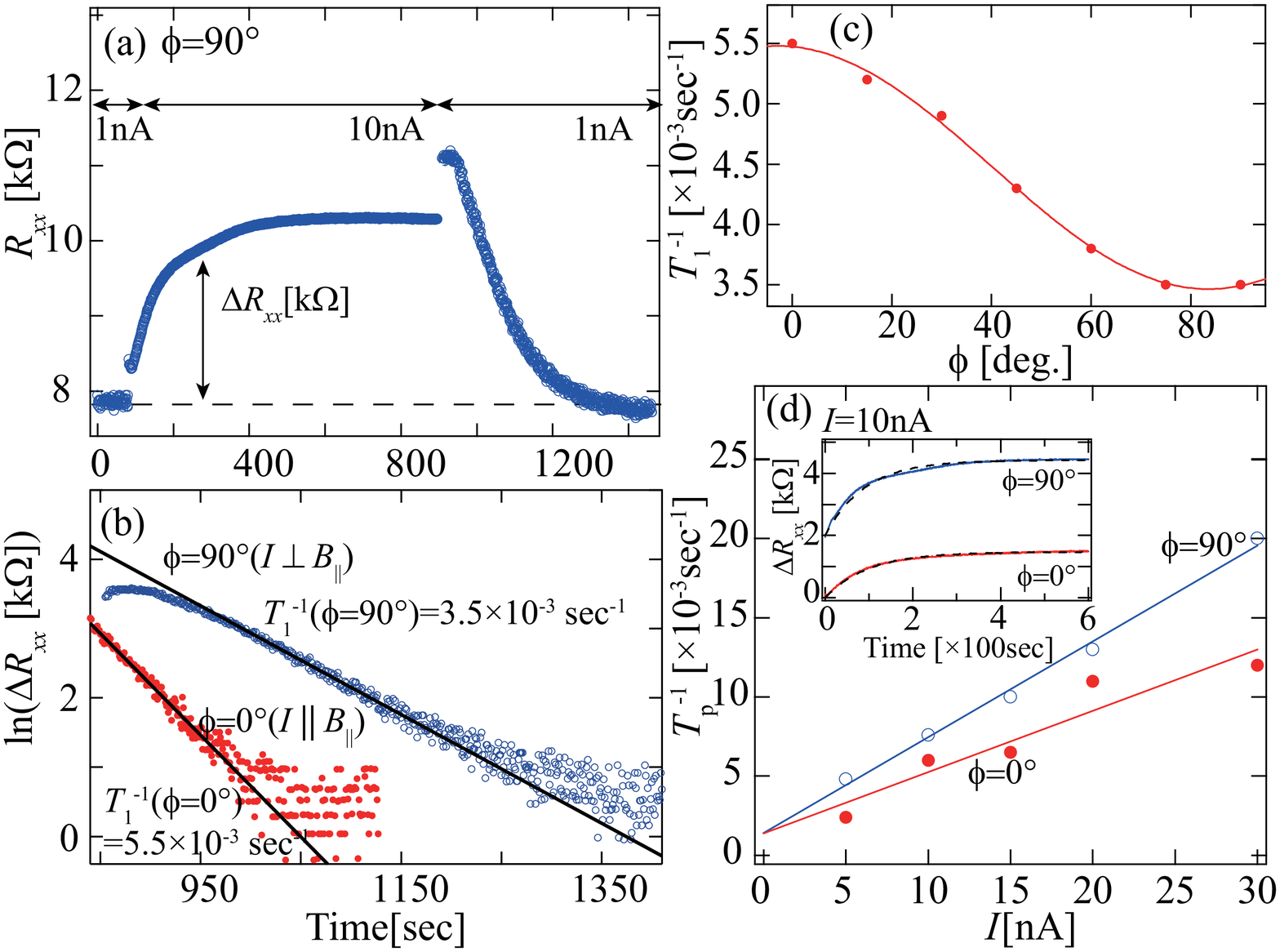}
\end{center}
  \caption{(Color online) 
(a) Relaxation measurement at the $R_{xx}$ peak in Fig.\,2 at $90^{\circ}$ at $T=70$\,mK.
(b) Log-scale plot of (a) after $I$ is reduced to 1\,nA.
Open and closed circles are for the data at $\phi=90^{\circ}$ and $0^{\circ}$, respectively.
Sold lines represent fitting by $\Delta R_{xx}$ $\propto$ exp($-t/T_{1}$).
(c) $T_{1}^{-1}$ as a function of $\phi$.
The curve is a guide for the eye.
$T_{1}^{-1}$ increases with decreasing $\phi$.
(d) $T_{\rm{p}}^{-1}$ as a function of pumping current.
Open (closed) circles are for the data at $\phi=90^{\circ}$ ($0^{\circ}$).
The lines are guides for the eye.
In the inset, the solid and dashed lines show the temporal evolution of $\Delta R_{xx}$ and the result of the fitting for $I=$10\,nA, respectively.
The traces for $\phi=90^\circ$ are vertically offset by 2\,k$\Omega$.
}
\end{figure}

By fitting the data with $\Delta R_{xx}$ $\propto$ exp($-t/T_{1}$), we obtain $T_{1}^{-1}=5.5\times10^{-3}$\,s$^{-1}$ for $\phi=0^\circ $ and $T_{1}^{-1}=3.5\times10^{-3}$\,s$^{-1}$ for $\phi =90^\circ $.
Since Hashimoto $et$\,$al.$ have shown that the magnitude of $\Delta R_{xx}$ is proportional to the nuclear spin polarization \cite{Hashimoto_PRL1} and the current pumping does not occur with $I=1$\,nA at $T = 70$\,mK, as we see in Fig.\,2, $T_{1}^{-1} $ reflects the relaxation of the current-induced nuclear spin polarization.
Figure 4 (c) displays $T_{1}^{-1}$ as a function of $\phi$ at the spin transition point.
The data show that $T_{1}^{-1}$ increases with decreasing $\phi$, as is consistent with the $\phi$ dependence of hysteresis.

We also analyze the nuclear spin pumping rate $T_{\rm{p}}^{-1}$ for several values of the pumping current.
By solving the rate equation d$P$/dt$=(1-P)\,T_{\rm{p}}^{-1}-P\,T_{1}^{-1}$, where $P$ is the nuclear spin polarization, we obtain the time evolution of $P$ as $P$ = $T_{\rm{p}}^{-1}/(T_{\rm{p}}^{-1}+T_{1}^{-1}$)\,[1$-$exp$\{-(T_{\rm{p}}^{-1}$+$T_{1}^{-1})$\,$t\}$].
By fitting the time evolution of $\Delta R_{xx}$ during the nuclear spin pumping process with $T_{1}^{-1}$ obtained by the relaxation measurements (Fig.\,4(b)), we can find $T_{\rm{p}}^{-1}$.
Figure\,4(d) shows $T_{\rm{p}}^{-1}$ as a function of pumping current for the two angles.
The inset in Fig.\,4(d) displays the temporal evolution of $\Delta R_{xx}$ at the peak with pumping current $I=10$\,nA for the two angles (solid lines), and the result of fitting (broken lines).
The data show that $T_{\rm{p}}^{-1}$ increases with the current value and has anisotropy.
At $I=30$\,nA, $T_{\rm{p}}^{-1} = 12\times10^{-3}$\,s$^{-1}$ for $\phi=0^\circ $ and $20 \times10^{-3}$\,s$^{-1}$ for $\phi=90^\circ $ are much larger than $T_{1}^{-1}$, reflecting that the hysteresis appears for both $\phi$s.
On the other hand, at $I=10$\,nA, while $T_{\rm{p}}^{-1}$ is about twice larger than $T_{1}^{-1}$ for $\phi=90^\circ$, the one for $\phi=0^\circ$ is similar to $T_{1}^{-1}$.
This difference leads to the anisotropy of hysteresis, as seen above.
Thus the anisotropy of hysteresis is relevant to the anisotropies of $T_{1}^{-1}$ and $T_{\rm{p}}^{-1}$.

Here, we discuss the origin of the anisotropies of $T_{1}^{-1}$ and $T_{\rm{p}}^{-1}$.
On the one hand, because electron and nuclear spins are coupled through the hyperfine interaction, it is known that the value of $T_{1}^{-1}$ is enhanced by low-frequency electron spin fluctuation \cite{Tycko_Science,Hashimoto_PRL1,Kumada_Science}.
Moreover, considering that a typical value of $T_{1}^{-1}$ in a quantum Hall system is about $10^{-3}$\,s$^{-1}$ \cite{Berg_PRL}, our result indicates that there are low-frequency electron spin fluctuations at the spin transition point, and the wavelength of the electron fluctuation has an anisotropy against $\phi$.
On the other hand, since the current-induced nuclear polarization is induced when an electron passes across the domain wall, the anisotropy of $T_{\rm{p}}^{-1}$ shows that the number of electrons passing across the domain wall is different between the two angles.
Note that, recently, Nakajima $et$\,$al.$ demonstrated the outward nuclear spin diffusion of a two-dimensional electron system and discussed that the decay of the nuclear polarization in the quantum well is influenced by not only electron-nuclear spin interactions but also the diffusion of nuclear spin polarization \cite{Nakajima_PRB}. In this experiment, the long pumping time period (800\,s) also leads to the diffusion of nuclear polarization to the outside of the quantum well. Although the nuclear spin diffusion affects $T_{1}^{-1}$, its influence should be independent of $\phi$, and the observed anisotropy of $T_{1}^{-1}$ is due to the anisotropy of electron-nuclear interactions.

Since there is an energy gap of electron spin fluctuations in the domains of the spin-polarized and -unpolarized QH states \cite{Kumada_Science}, the electron spin fluctuations hardly occur in each domain.
We speculate that the electron spins fluctuate at the domain walls and one possibility of the origin of the anisotropies is relevant to the morphology of the electronic domain structure.
If the domain structure has a unidirectional form determined by the direction of $I$, it will lead to a difference in the number of electrons passing across the domain wall and in the wavelength of electron spin fluctuation along the domain wall because the wavelength of the electron spin fluctuation in the domain wall may have a large anisotropy, i.e., the electron fluctuation has a long-wavelength mode along the domain wall, whereas across the domain wall, their wavelength is limited by the width of the domain wall. 
Previous works have shown that the occurrence of $R_{xx}$ at the phase transition point is caused by connecting the edge channels through the domain wall \cite{Kronmuller_PRL,Jungwirth_PRL_PRB}.
Moreover, the temporal evolution of $R_{xx}$ is induced by the modulation of the domain structure by the current-induced nuclear polarization \cite{Hashimoto_PRL1,Kraus_PRL}, implying that the domain wall extends in the across-the-width direction of the Hall bar.
Thus, it remains possible that the development of the domain structure is influenced by the current direction.

Finally, we must comment on the effect of $B_{\parallel}$ on the electron spin fluctuation in the domain wall.
Because the Zeeman energy tends to align electron spins along the applied magnetic field, the properties of electron spin fluctuation in the domain wall should be influenced by $B_{\parallel}$ and the direction of the fluctuation in the domain wall may be determined by the relative direction between $B_{\parallel}$ and the domain wall.

In summary, we studied the anisotropic behavior of the hysteretic transport of $R_{xx}$ induced by the current-induced nuclear spin polarization by altering the angles between $I$ and $B_{\parallel}$ at the spin transition point of the $\nu = 2/3$ fractional QH state.
We found that marked hysteresis appears when $B_{\parallel}$ is orthogonal to $I$, but it disappears when $B_{\parallel}$ is parallel to $I$, showing that the current-induced nuclear spin polarization decreases with $\phi$.
We also investigated $T_{1}^{-1}$ and $T_{\rm{p}}^{-1}$ at the spin transition point.
The $\phi$ dependence of $T_{1}^{-1}$ implies that there is a low-frequency electron spin fluctuation when the direction of $I$ is parallel to $B_{\parallel}$, and it is suppressed with increasing $\phi$.
In addition, the difference in $T_{\rm{p}}^{-1}$ indicates that the number of electrons passing across the domain wall depends on $\phi$.
These results lead us to the picture that the origin of the anisotropy of the current-induced nuclear spin polarization is related to the morphology of the electronic domain structure.
\\

\begin{flushleft}
{\bf Acknowlegements}
\end{flushleft}

We are grateful to T. Saku for growing the sample and K. Muraki, T. Satoh, and K. Hashimoto for useful discussion.
This work was supported in part by Grants-in-Aid for Scientific Research (Nos. 18740181, 20654028, 20029012, 21340082, 21540254 and 21740236) and a 21st Century COE Program Grant of the International COE of Exploring New Science Bridging Particle-Matter Hierarchy from the Ministry of Education, Culture, Sports, Science and Technology.
Y. Hirayama thanks ERATO Nuclear Spin Electronics Project.
K. Iwata thanks Research Fellowships of the Japan Society for the Promotion of Science for Young Scientists.



\begin{thebibliography}{9}
\bibitem{text}
See, for example, 
Z. F. Ezawa: {\it Quantum Hall Effect, Field Theoretical Approach and Related Topics} (World Scientific, Singapore, 2000). 
and 
S. M. Girvin and A. H. MacDonald: {\it Perspectives in Quantum Hall Effects} edited by A. Pinczuk and S. Das Sarma (Wiley, New York, 1997).

\bibitem{Eisenstein_etc_PRB}
J. P. Eisenstein, H. L. Stormer, L. N. Pfeiffer, and K. W. West: Phys. Rev. B {\bf 41} (1990) 7910;
L. W. Engel, S. W. Hwang, T. Sajoto, D. C. Tsui, and M. Shayegan: Phys. Rev. B {\bf 45} (1992) 3418;
W. Kang, J. B. Young, S. T. Hannahs, E. Palm, K. L. Campman, and A. C. Gossard: Phys. Rev. B {\bf 56} (1997) R12776.

\bibitem{Kukushkin_PRL}
I. V. Kukushkin, K. von Klitzing, and K. Eberl: Phys. Rev. B {\bf 55} (1997) 10607;Phys. Rev. Lett. {\bf 82} (1999) 3665.

%

\bibitem{Kumada_PRL_PRB}
N. Kumada, D. Terasawa, Y. Shimoda, H. Azuhata, A. Sawada, Z. F. Ezawa, K. Muraki, T. Saku, and Y. Hirayama: Phys. Rev. Lett. {\bf 89} (2002) 116802;
N. Kumada, D. Terasawa, M. Morino, K. Tagashira, A. Sawada, Z. F. Ezawa, K. Muraki, Y. Hirayama, and T. Saku: Phys. Rev. B {\bf 69} (2004) 155319.

\bibitem{Kronmuller_PRL}
S. Kronm$\ddot{\mathrm{u}}$ller, W. Dietsche, J. Weis K. von Klitzing, W. Wegscheider, and M. Bichler: Phys. Rev. Lett. {\bf 81} (1998) 2526;
S. Kronm$\ddot{\mathrm{u}}$ller, W. Dietsche, K. von Klitzing, G. Denninger, W. Wegscheider, and M. Bichler: Phys. Rev. Lett. {\bf 81} (1999) 4070.

\bibitem{Smet_Nature}
J.H. Smet, R.A. Deutschmann, W. Wegscheider, G. Abstreiter, and K. von Klitzing: Phys. Rev. Lett. {\bf 86} (2001) 2412;
J. H. Smet, R. A. Deutschmann, F. Ertl, W. Wegscheider, G. Abstreiter, and K. von Klitzing: Nature (London) {\bf 415} (2002) 281.

\bibitem{Hashimoto_PRL1}
K. Hashimoto, K. Muraki, T. Saku, and Y. Hirayama: Phys. Rev. Lett. {\bf 88} (2002) 176601.

\bibitem{Kraus_PRL}
S. Kraus, O. Stern, J. G. S. Lok, W. Dietsche1, K. von Klitzing, M. Bichler, D. Schuh, and W. Wegscheider: Phys. Rev. Lett. {\bf 89} (2002) 266801.

\bibitem{Stern_PRB}
O. Stern, N. Freytag, A. Fay, W. Dietsche, J. H. Smet, K. von Klitzing, D. Schuh, and W. Wegscheider:Phys. Rev. B {\bf 70} (2004) 075318.

\bibitem{Verdene_NaturePhys}
B. Verdene, J. Martin, G. Gamez, J. Smet, K. von Klitzing, D. Mahalu, D. Schuh, G. Abstreiter and A. Yacoby:Nat. Phys. {\bf 3} (2007) 392.






\bibitem{Jungwirth_PRL_PRB}
T. Jungwirth, S. P. Shukla, L. Smrcka, M. Shayegan, and A. H. MacDonald: Phys. Rev. Lett. {\bf 81} (1998) 2328;
T. Jungwirth and A. H. MacDonald: Phys. Rev. B {\bf 63} (2001) 035305.





\bibitem{Suzuki_cryogenics}
M. Suzuki, A. Sawada, A. Ishiguro, and K. Maruya: Cryogenics {\bf 37} (1997) 275.


\bibitem{Tycko_Science}
R. Tycko, S. E. Barrett, G. Dabbagh, L. N. Pfeiffer, and K. W. West: Science {\bf 268} (1995) 1460.

\bibitem{Kumada_Science}
N. Kumada, K. Muraki, and Y. Hirayama: Science {\bf 313} (2006) 329.

\bibitem{Berg_PRL}
A. Berg, M. Dobers, R. R. Gerhardts, and K. von Klitzing: Phys. Rev. Lett. {\bf 64} (1990) 2563.



\bibitem{Nakajima_PRB}
T. Nakajima, Y. Kobayashi, S. Komiyama, M. Tsuboi, and T.Machida: Phys. Rev. B {\bf 81} (2010) 085322.

\end{thebibliography}
\end{document}